\documentclass[12pt]{iopart}

\usepackage{graphicx}
\usepackage{color}
\begin{document}

\title[Quantum confinement effects in InAs-InP core-shell nanowires]{Quantum confinement effects in InAs-InP core-shell nanowires}

\author{Z Zanolli, M-E Pistol, L E Fr\"oberg and L Samuelson}

\address{Solid State Physics/The Nanometer Structure Consortium, 
Lund University, Box 118, S-221 00 Lund, Sweden}
\ead{zeila.zanolli@ftf.lth.se}
\begin{abstract}

\noindent
We report the detection of quantum confinement in single InAs-InP core-shell nanowires. The wires having an InAs core with $\sim 25$~nm diameter are characterized by emission spectra in which two peaks are identified under high excitation intensity conditions. The peaks are caused by emission from the ground and excited quantized levels, due to the quantum confinement in the plane perpendicular to the nanowire axis. We have identified in the emission spectra different energy contributions   related to the wurtzite structure of the wires, the strain between the wurtzite core and shell, and the confinement energy of the InAs core. Calculations based on 6-band strain-dependent ${\bf k\cdot p}$ theory allow the theoretical estimation of the confined energy states in such materials and we found these results to be in good agreement with those from the photoluminescence studies.
\end{abstract}

% \maketitle

\section{Introduction}
The investigation of the properties of nanometer-size materials is an essential step towards a better understanding of the fundamental physics underlying their behavior and hence to further developments in nanotechnologies. Semiconductor nanowires (NWs) have the double feature of providing a tool for the study of the physical properties of low-dimensional systems and, at the same time, to provide the building blocks of electronic and photonic nanoscale devices \cite{Hu1999}.  Nanoscale field-effect transistors \cite{ Bryllert2006a, Bryllert2006b, Duan2001, Cui2001, YHuang2001}, inverters \cite{Cui2001, YHuang2001}, and logic gates \cite{YHuang2001} are some examples of NW-based devices. One-dimensional ($1$D) electronics can also take advantage of the growth of heterostructures along the NW axis, as has been demonstrated in resonant tunneling diodes \cite{Bjork2002} and single-electron transistors \cite{Thelander2003}. Concerning photonic applications, devices such as LEDs \cite{Duan2001}, photodetectors \cite{Wang2001}, and lasers \cite{MHHuang2001, Johnson2001} based on NWs have been reported. The results achieved so far are encouraging to pursue developments in the bottom-up fabrication approach and, at the same time, to call for investigations of the electrical and optical properties of NWs to define  the conditions for future applications.

The NW structure is a favorable environment to achieve 1D systems, {\it i.e.} to force the motion of the electrons along only one direction, the NW axis. Indeed, if the NW diameter is small enough, the electrons will be confined by a potential well in the radial direction with a consequent energy quantization in that plane. Hence NWs exhibit physical properties which are intrinsically different from those of quantum dots and bulk materials. The Bohr radius of the exciton in a bulk crystal \cite{Yoffe1993, Ascrofh-Mermin} provides the length scale for the onset of quantum-confinement effects. This implies that in the nanowires presented here quantum confinement  can be observed when the core diameter is smaller than 35~nm, the exciton Bohr radius in bulk InAs. When this condition is met, we expect the formation of quantized levels in both conduction and valence bands of the semiconductor. Analogous to what happens in a quantum well, the bandgap in a quantum wire is larger than the gap of the bulk material. Moreover, transitions between the localized quantum levels now become possible.
%Since for both bands the quantum ground state levels are at higher energy than in the bulk material, the resulting bandgap is larger than the bulk one. Besides, transitions between the localized levels became now possible.

Electron quantum confinement was demonstrated via photoluminescence (PL) measurements in GaAs \cite{Duan2000} and InP \cite{Gudiksen2002, Bhunia2003} nanowires as a blue-shift of the emission energy with decreasing diameter of the wires, and via scanning tunneling microscopy (STM) in Si nanowires \cite{Ma2003}.~In InAs nanowires with diameter smaller than 30~nm the observed drop in the wire conductance was explained as an effect of the quantum confinement \cite{Thelander2004}. We report here studies of quantum confinement effects in InAs-based NWs. The wires were grown via Chemical Beam Epitaxy (CBE) and the optical characterization was performed via low temperature micro-photoluminescence measurements ($\mu$-PL) on single NWs. Due to the low emission efficiency of  un-capped InAs NWs, we chose to study InAs-InP core-shell structures  similar to those reported in \cite{Zanolli2006}. The shell material provides a passivation of the core surface, hindering the non-radiative electron-hole recombination through surface states, a mechanism that limits the emission efficiency of the core. 

After taking into account the shift towards higher energies due to the strain and to the wurtzite structure of the InAs NWs, we have found that the PL peaks are further blue-shifted with respect to the bulk energy gap of the zinc-blende phase due to quantum confinement. Besides that, as a clear signature of level quantization, we report the observation of the first excited state above the gap.   We compare the experimental results to accurate calculations based on 6-band strain-dependent ${\bf k\cdot p}$ theory in the so called envelope function approximation \cite{Gershoni1993} and we find good agreement between experiments and theory.
%When compared, the experimental results were found to be in agreement to accurate calculations based on 6-band strain-dependent ${\bf k\cdot p}$ theory in the so called envelope function approximation \cite{Gershoni1993}.

\section{Methods of investigation}

\subsection{Sample fabrication}
\begin{figure}[htbp]
\begin{center}
\includegraphics[width=10cm]{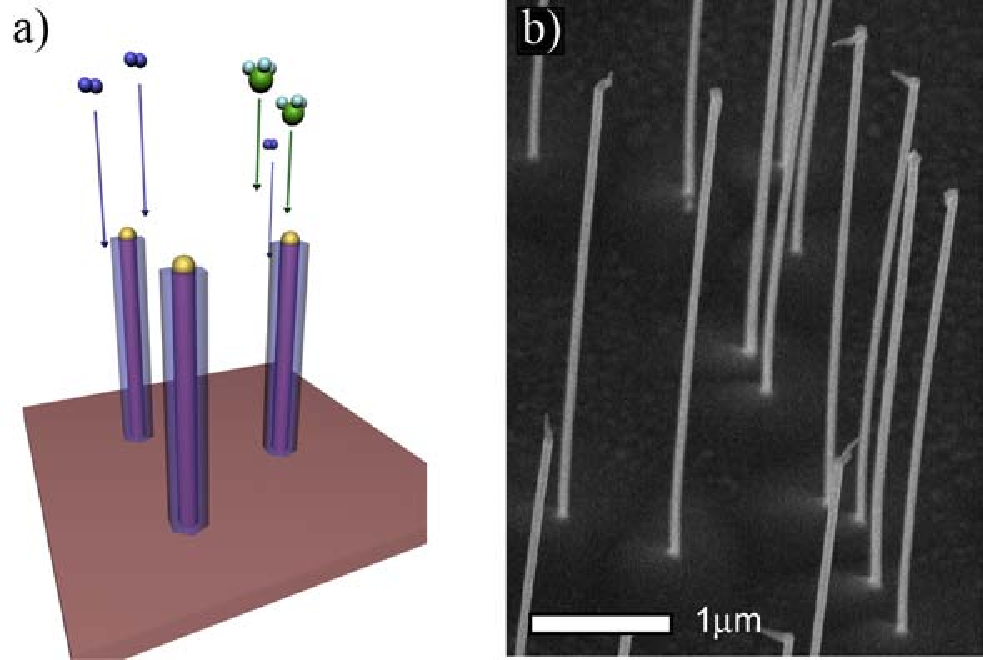}
\caption{(a) Schematic representation of the growth process of InAs-InP core-shell NWs  and (b) SEM image of the as-grown wires (25~nm core diameter, 20~nm shell thickness).}
\label{growthSEM}
\end{center}
\end{figure}

The nanowires are grown using CBE, which is a high vacuum growth technique similar to Molecular Beam Epitaxy, but using organometallic source molecules. The material is supplied in a beam directed towards the sample \cite{Ohlsson2001, Jensen2004} and  -- due to the low pressure  -- the mean free path of the material is much longer than the chamber size. The first step of the sample fabrication is the deposition of Au aerosol particles on a InAs(111)B substrate.
%, then the sample is loaded into the growth chamber and the actual nanowires formation takes place. 
The aerosol particles are produced in a home built system \cite{Magnusson1999}, where Au is evaporated, condensed and size selected.  The size selection is important since it determines the diameter of the nanowire. In this study we used Au particles 20~nm and 40~nm wide, leading to  $\sim25$~nm and $\sim45$~nm core diameters, respectively. The Au particles enhance the growth rate beneath them, resulting in rod like structures on the surface with the Au particle sitting on top, schematically shown in figure \ref{growthSEM}a. The most referenced growth model for nanowires is based on the Vapor-Liquid-Solid (VLS) mechanism \cite{Wagner}, in which material is supplied in the vapor phase which preferably condenses on the liquid metal-semiconductor alloy. This creates a stronger driving force for growth at the liquid-solid interface compared to the vapor-solid interface and growth material precipitates out into the crystal from the alloy. 
%However, we believe that the nanowires in our system grow from a solid particle because the growth temperature is lower than any melting point in the phase diagram . 
However, since the growth temperature is lower than any melting point in the phase diagram, the nanowires in our system grow from a solid particle. This has led to the suggestion of Vapor-Solid-Solid (VSS) growth \cite{Persson2004, Dick2005}.

We have grown InAs wires at 425~$^{\rm o}$C, under condictions resulting in a wurzite crystal structure of the core, and then lowered the temperature to 370~$^{\rm o}$C and switched precursors to allow for radial (20~nm thick) InP growth as in \cite{Zanolli2006}. A Scanning Electron Microscope (SEM) image of such a sample can be seen in figure \ref{growthSEM}b at an angle of 30$^{\rm o}$. The sources used are Tertiarybutylarsine (TBAs), Tertiarybutylphosphine (TBP) and Trimethylindium (TMIn) for As, P and In, respectively, and the corresponding pressures in the gas lines to the growth chamber are 1.5~mbar, 3.0~mbar and 0.15~mbar. The TBAs and TBP are thermally cracked upon entering the growth chamber while the TMIn decomposes on the substrate surface. During the shell growth, the axial nanowire growth is not fully suppressed and  InP grows under the Au particle under non-ideal conditions, as can be seen in figure \ref{growthSEM}b as an extra feature at the top of the wires.

\subsection{PL Measurements on single NWs}
After growth, the NWs were mechanically transferred onto a gold patterned Si/SiO$_2$ substrate where single wires could be located and their optical properties studies on a single NW level. The substrate with the wires was inspected using an optical microscope ($100\times$ objective, dark field) and -- after PL measurements --  a SEM to identify the single wires more suitable for the optical characterization, {\it i.e.} those wires which are well isolated and far from any other particle eventually present on the substrate by more than $\sim20 \mu$m. 
The spectra from single NWs were obtained using a $\mu$-PL setup optimized for detection of radiation in the near IR ($0.9 \mu$m -- $2.0 \mu$m). The NWs were excited with the 532~nm line of a frequency doubled Nd-YAG laser. The laser light was focused onto the sample, mounted on the cold finger of a continuous flow helium cryostat for low temperature ($\sim 5$ K) measurements. The light emitted from the NW was collected by a long working distance reflective objective (NA = 0.28) of a microscope and was focused onto the entrance slit of a spectrometer. Then the emitted light was imaged or spectrally dispersed on the HgCdTe focal plane array of a liquid N$_2$ cooled camera.

\subsection{Calculations}\label{calculations}
For the interpretation of the experimental results, we initially computed the strain tensor profile of the NWs using linear continuum elasticity theory. Our calculation is fully three-dimensional and was done on a $100\times100\times100$ grid where the strain energy was minimized by the conjugate gradient method. Using the so-obtained strain tensor element we computed the electronic states within the so called ${\bf k\cdot p}$ theory allowing mixing among six bands in the valence band \cite{Prior1998}. The conduction band was treated in the single-band approximation. We included the piezoelectric polarization in the calculation. Since the structure is a wire we have a continuum of states and it was necessary to compute a large number of eigenvalues in order to identify the radially excited states. The differential system was discretized on a grid and formulated as a finite difference problem. In order to get reasonable computing times we truncated the grid from $100\times100\times100$ to $60\times60\times60$ when doing the electronic structure calculation. In contrast to the power-law decay of the strain, the wave-functions have an exponential decay in the barrier and it is thus appropriate to use a smaller box for the electronic calculations than for the strain calculations. The Lanczos algorithm was used to fit the eigenvalues.

\section{Results and discussion}

\begin{figure}[htbp]
\begin{center}
\includegraphics[width=14cm]{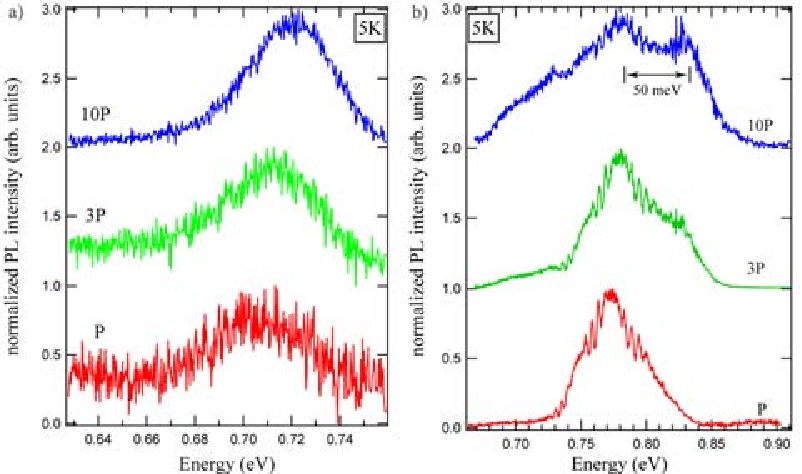}
\caption{PL emission spectra at different excitation laser densities from 45~nm--20~nm (a) and 25~nm--20~nm (b) core-shell InAs-InP single NWs. In the 25~nm core wire the two peaks from the ground state and first excited quantized states are visible.}
\label{PL}
\end{center}
\end{figure}

To find evidence of quantum confinement effects in InAs NWs we considered here two sets of NWs differing in their diameter, {\it i.e.} 
25~nm and 45~nm in diameter.
%NWs grown using 20~nm and 40~nm wide Au seed particles, respectively. Because of the limitations in emission efficiency of the bare nanowires, we decided to cap them with an InP shell, as described in \cite{Zanolli2006}.
The PL spectra from such wires are shown in figure \ref{PL}a and \ref{PL}b for 45~nm and 25~nm core width, respectively. At low excitation power density, a single peak is detected at about 771~meV (FWHM = 50~meV) and 703 meV (FWHM = 60~meV), respectively. Evidence of state filling is visible in both spectra as an energy blue shift of the main peak with the increase of the excitation power density \cite{Castrillo1995}. The striking difference between the two samples is that in the emission from the 25~nm wire at high excitation power density two peaks are visible at energies of about 781~meV and 831~meV. This can be interpreted as evidence of energy quantization in the radial direction, where the low and high energy peak can be interpreted as the ground and first excited quantized levels, as we will discuss in this section. 

In the analysis of these emission spectra one should consider the fact that there are different effects contributing to the blue shift of the wire emission compared to bulk InAs zinc-blende. First of all, the shell thicknesses of both NWs were 20~nm, so the energy shifts due to the strain is larger for the small core diameter wire, resulting into a higher emission energy.
% between the core and shell material will be different for the small and large core diameter wire, resulting into a higher emission energy for the small core wires.

\begin{figure}[htbp]
\begin{center}
\includegraphics[width=12cm]{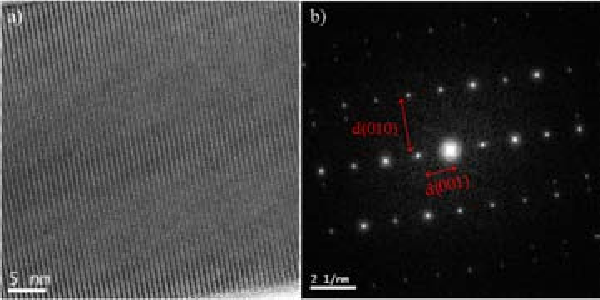}
\caption{TEM image (a) and its Fourier transform (b) of an InAs wire ($\sim 42$~nm wide). The wurtzite lattice constants are $a_{wz} = 2 {\rm d}(010)/\sqrt{3}$ and $c = {\rm d}(001)$. {\it (Courtesy of Jacob B. Wagner)}}
\label{TEM}
\end{center}
\end{figure}

We should also take into account the fact that these NWs have wurtzite ({\it wz}) crystal structure, 
%instead of zinc-blende ({\it zb}), as it is the case for the bulk InAs and InP compounds. 
while zinc-blende ({\it zb}) is the crystal structure observed for bulk InAs and InP compounds. 
The wurtzite phase is demonstrated by the TEM analysis performed on such wires. Figure \ref{TEM}a shows the TEM image of a bare InAs NW together with the Fourier transform of the image, from which the lattice constants are measured as $a_{wz} = 4.2839$~{\AA} and $c = 6.9954$~{\AA}. This fact affects the energy gap of both materials, so that the {\it wz} phases have a larger gap relative to the {\it zb} phases as theoretical studies based on Density Functional Theory (DFT) reveal \cite{Yeh1994, Murayama1994}. We should mention that the energy gap predicted via DFT is usually underestimated because it is an excited state property and can be correctly described using Many-Body Perturbation Theory. This treatment is beyond the purview of this article and the reader is referred to the work done by one of the authors on the calculation of the quasiparticle band structure in the so-called dynamically screened exchange approximation (GWA) \cite{Hedin1965, Hedin1969} of InAs in the wurtzite phase \cite{Zanolli2006b}. From this study it has been found that the difference in the energy gaps of InAs {\it wz} and {\it zb} is $\Delta{gap} =  55.7$~meV, leading to $470.7$~meV as energy gap of the InAs in the wurtzite phase at $0$~K.

In the case of the InP compound, the energy gap of the wurtzite phase has been measured in $\sim50$~nm thick nanowires to be $\sim80$~meV higher than the one of the corresponding zinc-blende phase \cite{Mattila2006}.
%Since the increase in energy gap due to the {\it wz} phase is higher in the InP than in the InAs material system and due to the fact that higher energy gap corresponds to a smaller {\it ``a"} lattice constant, it follows that the difference between the {\it ``a"} lattice constant of the two materials in the {\it wz} phase is larger with respect to the {\it zb} case. 
Since the increase in energy gap due to the {\it wz} phase is higher in the InP than in the InAs material system, it follows\footnote{In tight binding approximation it can be shown that in direct band-gap III-V semiconductors a higher energy gap corresponds to a smaller {\it ``a"} lattice constant.} that the difference between the {\it ``a"} lattice constant of the two materials in the {\it wz} phase is larger as compared to the {\it zb} case.  Hence the lattice mismatch between the two materials is higher when they are in the {\it wz} phase and, consequently, the blue shift in energy due to strain will be higher when InAs and InP are in the wurtzite phase. We denote by  $\Delta{strain} = E_{strainWZ} - E_{strainZB}$ the increase in the InAs emission energy due to this effect.
We can summarize all the mentioned contributions to the InAs emission as following

\begin{equation}
E = E_{bulkWZ} + E_{strainWZ} + E_{QC} ~,
\end{equation}	
where $E$ is the measured emission energy, $E_{bulkWZ}$ is the band-gap of the bulk InAs in the {\it wz} phase  and $E_{QC}$ is the contribution given by quantum confinement.

From the data and the modelling  reported in \cite{Zanolli2006} it is possible to evaluate the difference in the confinement energy between the 25 nm and 45 nm core wires $\Delta E_{QC} = E_{QC}(25~{\rm nm}) - E_{QC}(45~{\rm nm})$. Indeed, for each diameter, the dependence of the measured emission energy $E$ and that one calculated using 8-band strain-dependent ${\bf k\cdot p}$ theory without the inclusion of quantum confinement effects for the {\it zb} phase ($E_{calc} = E_{bulkZB} + E_{strainZB}$) can be written as

\begin{equation}
\label{Ediff}
E - E_{calc} = \Delta{strain}  + \Delta{gap} + E_{QC} ~,
\end{equation}
where we use the measured emission energies at low excitation power (771 meV and 703 meV) and the values obtained from an exponential fit of the calculated data, {\it i.e.} $567$ meV and $537$ meV for the 25~nm--20~nm and 45~nm--20~nm core-shell wires, respectively. Writing equation (\ref{Ediff}) for both (25 nm and 45 nm) core diameters and taking their difference leads to $\Delta E_{QC} = 38$ meV. It is worth noting that this value is obtained without any assumption on $\Delta strain$ and $\Delta gap$. 

\begin{figure}[htbp]
\begin{center}
\includegraphics[width=10cm]{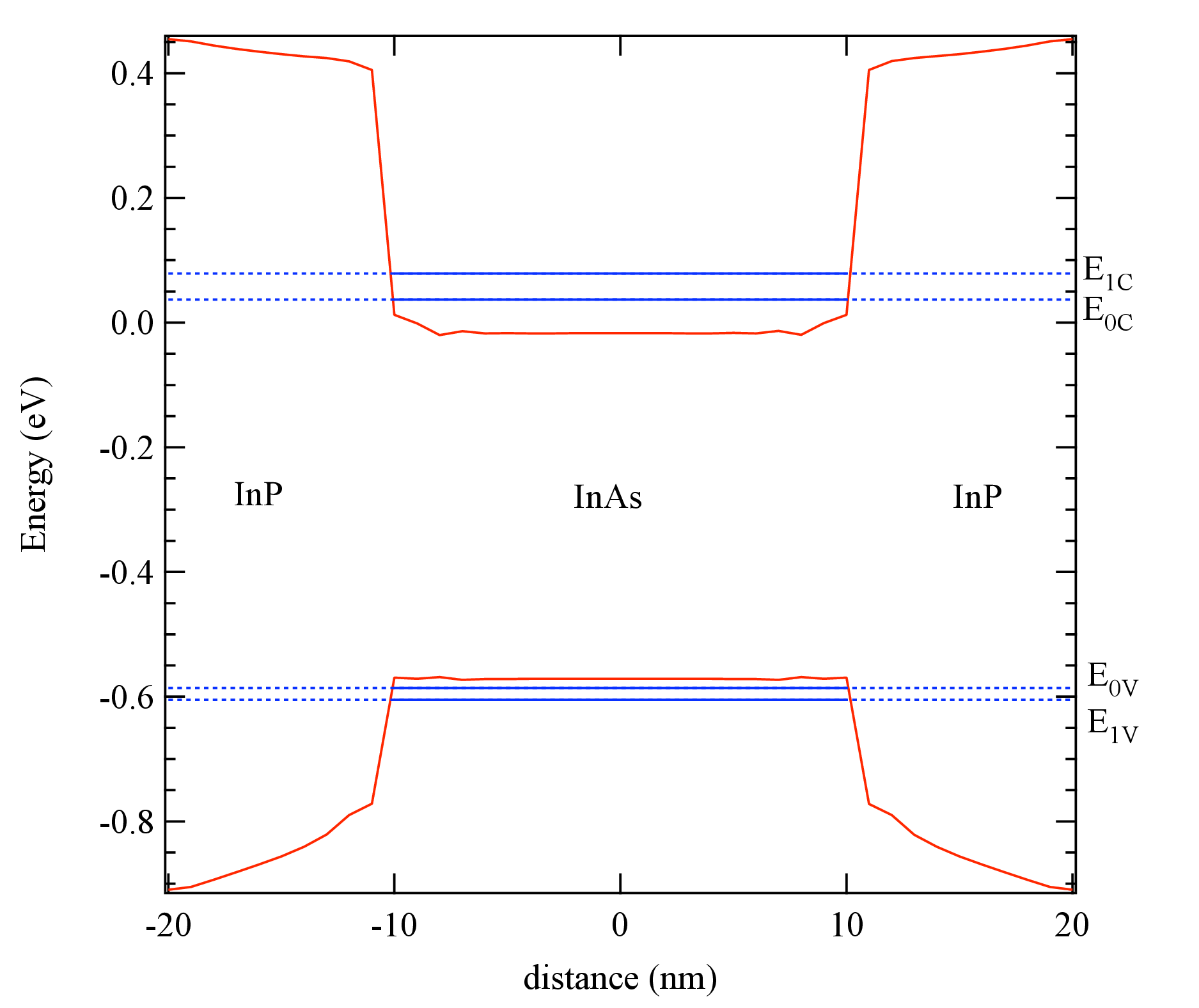}
\caption{Location of the quantized energy levels with respect to the band structure. The discrete states are calculated for an InAs-InP 25 nm - 20 nm core-shell wire.}
\label{Energy_bands_levels}
\end{center}
\end{figure}

Calculations of the confinement energies and of the energies of the ground ($E_{0c}, E_{0v}$) and first ($E_{1c}, E_{1v}$) excited states in InAs for both electrons and holes in the InAs/InP wire system were performed according to the method outlined in section \ref{calculations}. These calculations, whose results are summarized in table \ref{conti} and are visualized in figure \ref{Energy_bands_levels}, were done for wires having InAs cores 25~nm and 45~nm wide, which is the expected average diameter when the seed Au particle is 20~nm and 40~nm, respectively. The so-obtained confinement energy of both electrons and holes leads to $E_{QC}(45~{\rm nm}) = 24.7$ meV and $E_{QC}(25~{\rm nm}) = 68.7$ meV, hence $\Delta E_{QC calc} = 44$ meV. Considering the uncertainties in the determination of the actual core diameter, we find that the results agree. 

\begin{table}[htbp]
\caption{\label{conti}The confinement energies and the quantized energy levels for the InAs wires under study, with the inclusion of the strain effects due to the InP shell in zb materials. The energies are reported in meV, the lengths in nm.}
\begin{indented}
\item[]\begin{tabular}{@{}lllllllll}
\br
Au diam  & InAs diam & $E_{QC}{\rm e}^-$ & $E_{QC}{\rm h}^+$ & $E_{QC}$ & $E_{0c}$ & $E_{0v}$ &  $E_{1c}$ & $E_{1v}$ \\
\mr
20 & 25 &  54 & 14.7 & 68.7 & 37 & -586 & 79 & -605 \\
40 & 45 &  14 & 10.7 & 24.7 &  -3 & -582 & 16 & -587\\
\br
\end{tabular}
\end{indented}
\end{table}

We then compared the calculated transition energies with the measured ones in the 25 nm wires at high excitation power density. The ground and first excited state emission energies are calculated as $E_0 = E_{0c} - E_{0v} = 623$~meV and $ E_1 = E_{1c} - E_{1v} = 684$~meV, respectively. These values have to be compared with the first and second peak in the measured spectrum at 781 meV and 831 meV. At first we notice that the energy difference between these peaks is about 50 meV and the calculated one is 61 meV. Hence the energy separation of the two peaks is reproduced well (within $\sim10$ meV) by our theoretical model.
Moreover, since the calculations are performed for {\it zb} InAs and InP materials, the difference between the calculated and measured (at low excitation power density) values is given by $\Delta{gap}$ + $\Delta{strain}$, that therefore amounts to

\begin{equation}
\Delta gap + \Delta strain = 148 ~{\rm meV}~.
\end{equation}

If we then use the GW value for the difference in energy gaps ($\Delta{gap} = 55.7$~meV), we obtain $\Delta{strain} = 92.3$~meV.

\section{Conclusion}
In this paper we have reported the optical study and modeling of quantum confinement effects in the emission from the InAs core of InAs/InP core-shell strained nanowires. The PL measurements performed on single NW having small ($\sim25$~nm) core diameter are characterized by a double peak emission due to the formation of quantized energy levels in the band structure. The energy separation between the fundamental and excited peak of about 50~meV is well reproduced by the theoretical calculations. Since the wires have the wurtzite crystal structure, while the calculations describe zinc-blende material systems, we have taken into account the increase in the actual energy gap and the difference in the strain between the wurtzite materials. Each of these two effects causes a blue shift of the InAs emission to a total contribution of 148~meV. Beside this increase in emission energy, the observed fundamental peak is further shifted due to quantum confinement, consistently with the calculations. Finally, we compared the experimental and calculated energy difference between the confinement energy in 25~nm and 45~nm wires, finding that the two estimations are in agreement.

\section*{Acknowledgments}
This research was conducted within the nanometer structure consortium in Lund and supported by the European CommunityÕs Human Potential Program under contract HPRN-CT-2002-00298, the EU program NODE 015783,  the Swedish Foundation for Strategic Research (SSF), the Swedish Research Council (VR) and Knut and Alice Wallenberg Foundation. 
Dr. J. B. Wagner is gratefully acknowledged for TEM imaging.

\end{document}